\documentclass[conference]{IEEEtran}
\usepackage[utf8]{inputenc}
\usepackage{geometry}
\usepackage{amsmath, amsthm, amssymb}
\usepackage{mathtools}
\usepackage{cases} %%%for some reason this has to be after amsmath%%%
\usepackage{braket}
\usepackage{upgreek} %nicer Pi symbols
\DeclareMathAlphabet{\mathpzc}{OT1}{pzc}{m}{it}% imports the script for the hilbert space dimension

\usepackage[numbers]{natbib}

\usepackage{thmtools}
\usepackage{thm-restate}

\usepackage{hyperref}
\usepackage{float}

\usepackage{graphicx}
\usepackage{adjustbox,lipsum}
\usepackage{xcolor}
\usepackage{tikz}
\usepackage{rotating}

\usepackage{algorithm}
\usepackage{algpseudocode}

\usepackage{pdflscape}

\newtheorem*{remark}{Remark}
\newtheorem{lemma}{Lemma}
\newtheorem{definition}{Definition}

\newtheorem{problem}{Problem}
\usepackage{tikz}
\usetikzlibrary{quantikz}
\usetikzlibrary{decorations.text}
\usetikzlibrary{decorations.pathreplacing}

\newcommand{\cheby}{\mathrm{T}}
\newcommand{\chebyt}{\mathrm{U}}
\newcommand{\erf}{\mathrm{erf}}
\newcommand{\sgn}{\mathrm{sign}}
\newcommand{\dom}[1]{\mathcal{D}_{1}}

\newcommand{\linfnorm}[1]{\left|\left|#1\right|\right|_{\infty}}

\usepackage{hyperref}
\usepackage{cleveref}
\crefname{problem}{problem}{problem}

\title{Mostly Harmless Methods for QSP-Processing with Laurent Polynomials}

\author{\IEEEauthorblockN{S. E. Skelton}
\IEEEauthorblockA{Quantum Information Group\\Institute For Theoretical Physics\\University of Leibniz Hannover\\
Hannover, Germany\\
Email: shawn-skelton@itp.uni-hannover.de}
}

\begin{document}
\maketitle
\begin{abstract}
   Quantum signal processing (QSP) and its extensions are increasingly popular frameworks for developing quantum algorithms. Yet QSP implementations still struggle to complete a classical pre-processing step ('QSP-processing') that determines the set of $SU(2)$ rotation matrices defining the QSP circuit. We introduce a method of QSP-processing for complex polynomials that identifies a solution without optimization or root-finding and verify the success of our methods with polynomials characterized by floating point precision coefficients. We demonstrate the success of our technique for relevant target polynomials and precision regimes, including the Jacobi-Anger expansion used in QSP Hamiltonian Simulation. For popular choices of sign and inverse function approximations, we characterize regimes where all known QSP-processing methods should be expected to struggle without arbitrary precision arithmetic.
\end{abstract}

%%%%%%%%%%%MAIN SECTION%%%%%%%%%%%
\section{Introduction}
At its core, QSP/QSVT is simply a routine for computing a polynomial of the eigenvalues/singular values of a matrix on a quantum circuit. The polynomial is usually an approximation of some target function $f_{targ}$, and then the polynomial degree is a function of the precision of that approximation. The techniques are appealing for their conceptual simplicity \cite{martyn_grand_nodate} and competitive cost scaling \cite{martyn_grand_nodate, rall_faster_2021}; an increasing literature of QSP/QSVT applications, extensions, and modifications testifies to this heuristic's increasing popularity. Because the QSP heuristic often allows one to obtain an extremely short circuit at the cost of low precision, some QSP routines have even been proposed for near-term benchmarking \cite{dong_quantum_2021, dong_random_2021} or as intermediate quantum algorithms \cite{dong_ground_2022}.

QSP initially struggled to gain traction due to difficulty solving a classical pre-processing problem \cite{childs2018towards} we call 'QSP-processing'. To specify a QSP circuit, users must formulate their problem as a function on the eigenvalues $z\in U(1)$ or singular values $\xi\in [0, 1]$ of some matrix $A$, accessible through unitary oracle $U_A$. The user must then design and bound a 'suitable'\footnote{'Suitable' is convention dependent, and in our case defined precisely in \cref{lemma:QSP_Haah}.} polynomial  $P$ approximating that function, and then solving for the set of $SU(2)$ rotations that define the QSP/QSVT circuit. 

A swath of advances in QSP-processing \cite{chao2020finding, Haah2019product, dong_efficient_2021, Ying2022stablefactorization} have made the problem tractable for real-valued functions implemented in the most common conventions of QSP. In the now standard approach which extends to QSVT  \cite{gilyen_quantum_2019, martyn_grand_nodate, lin_lecture_2022}, QSP is a procedure mapping a  $x\in[-1, 1]$ to a bounded, definite parity, complex polynomials in $x$. Very often, these are represented as a linear combination of Chebyshev polynomials. We will call these conventions collectively 'angle-QSP'. One can then construct an arbitrary complex polynomial on variable $x=\cos\theta$ on a quantum circuit, using a linear combination of unitaries (LCU) \cite{berry_hamiltonian_2015} to add four angle-QSP sequences on the circuit \cite{gilyen_quantum_2019}.

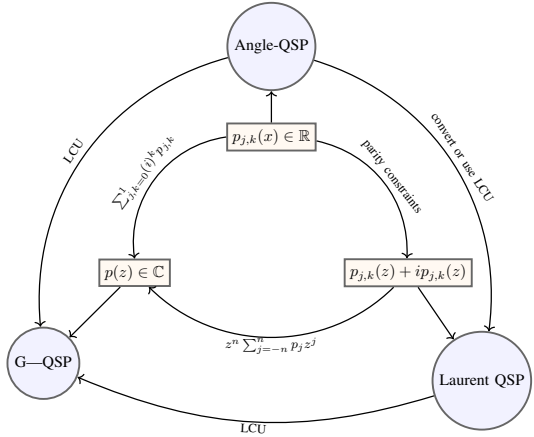
\begin{figure}[h]
    \centering
\scalebox{0.6}{
      \begin{tikzpicture}[
roundnode/.style={circle, draw=black!60, fill=blue!5, very thick, minimum size=7mm},
squarednode/.style={rectangle, draw=black!60, fill=orange!5, very thick, minimum size=5mm}, 
]
    \node[squarednode] (one) at (0,3)   {$p_{j,k}(x)\in \mathbb{R}$}; 
    \node[squarednode] (two) at  (3,0) {$p_{j,k}(z)+ip_{j,k}(z)$};
    \node[squarednode] (three) at (-3,0) {$p(z) \in \mathbb{C}$};
    \draw [->,thick,postaction={decorate,decoration={raise=1ex,text along path,text align=center, text={|\footnotesize|{parity constraints}{}}}}] (one) to [bend left=40] (two);
    \draw [<-,thick,postaction={decorate,decoration={raise=1.5ex,text along path,text align=center,text={|\footnotesize|{$\sum_{j, k=0}^1 (i)^{k}p_{j, k}$}{}}}}] (three) to [bend left=45] (one);
    \draw [<-,thick,postaction={decorate,decoration={raise=-1.5ex,text along path,text align=center,text={|\footnotesize|{$z^n\sum_{j=-n}^n p_jz^j$}{}}}}] (three) to [bend right=45] (two);
    \node[roundnode] (four) at (-5,-2)   {G—QSP}; 
    \node[roundnode]  (five) at (5,-2, )  {Laurent QSP};
    \node[roundnode]  (six) at (0, 5)  {Angle-QSP};
    \draw [<-,thick,postaction={decorate,decoration={raise=-1.5ex,text along path,text align=center,text={|\footnotesize|{LCU}{}}}}] (four) to [bend left=-20] (five);
    \draw [->,thick,postaction={decorate,decoration={raise=1ex,text along path,text align=center,text={|\footnotesize|{convert or use LCU}{}}}}] (six) to [bend left=40] (five);
    \draw [<-,thick,postaction={decorate,decoration={raise=1ex,text along path,text align=center,text={|\footnotesize|{LCU}{}}}}] (four) to [bend left=40] (six);
    \draw [->,thick,] (one) to (six);
    \draw [->,thick] (two) to  (five);
    \draw [->,thick,] (three) to (four);
    \end{tikzpicture}  
    } 
    \caption{Commonly used polynomials for QSP and associated QSP-conventions, assuming all polynomials obey $\linfnorm{p}\leq 1$. While some complex polynomials can be solved with angle-QSP, the convention is predominantly used for real ($k=0$) or purely imaginary ($k=1$) polynomials with definite parity ($j=0, 1$). Coordinate transformation $x=\cos\theta\rightarrow z^{i\theta}$ allows polynomials of the form $p_{0, j}(z)+p_{1, (j+1)\mod 2}(z)$ to be solved with Laurent-QSP. Complex polynomials (or Laurent polynomials) can be constructed as combinations of Laurent or real polynomials fitting the parity restrictions above. In some cases, Laurent and Angle QSP circuits are directly convertible.  Linear combination of unitaries (LCU) techniques can be used on Angle or Laurent-QSP circuits to obtain more general polynomials. }
    \label{fig:convention_sketch}
\end{figure}

Yet it is conceptually just as natural to treat QSP as a series of controlled $U, U^{\dag}$ gates, where the control basis changes with each iteration. We call this 'Laurent-QSP', after the type of polynomial it builds on the circuit. Laurent-QSP gives a circuit with a well-understood specialization to angle-QSP \cite{Haah2019product, martyn_grand_nodate, wang_energy_2022}.  One maps $U$'s eigenvalues $z\in U(1)$ to a bounded Laurent polynomial, and then angle-QSP emerges as a special case where the circuit can be rewritten as a product of Pauli rotations. Quite recently QSP circuits for more general polynomials have been introduced \cite{motlagh_generalized_2023, yamamoto2024robust}. One of these, "G-QSP" was introduced alongside a QSP-processing strategy which is successful for large random polynomials \cite{motlagh_generalized_2023}. Although similar to Laurent-QSP, these procedures relax the requirements on target polynomials. However, the performance of these more general techniques on analytically constructed polynomials of practical interest is unknown. We summarize three common approaches to building polynomials with QSP, the available QSP-conventions for each, and the connections between each convention in \cref{fig:convention_sketch}. 

Our main contribution is a QSP-processing algorithm for Laurent polynomials. Drawing heavily from the QSP-processing strategy introduced by \cite{Haah2019product}, we replace a polynomial root-finding subroutine, which is one of the major bottlenecks of the process. Instead, we identify the polynomial decomposition usually solved through root-finding as a Fejer problem in complex analysis, and compute solutions using Newton-Raphson iteration. 

Of the dozens of recent applications of QSP/QSVT, only a few analytical polynomials are given in the literature. However, most existing benchmarking results either use random polynomials \cite{motlagh_generalized_2023}, or identify polynomials using Remez optimization methods \cite{dong_efficient_2021}. Thus there exists a conceptual gap between the quantum algorithms literature, which uses analytical expansions to construct QSP-algorithms with bounded precision to the desired function, and QSP-applications literature, which relies heavily on optimization strategies to find solutions. Even then, practical applications to date have accepted extremely low precision \cite{lapworth2024evaluation}, or integrated trotter decomposition steps \cite{novikau_quantum_2022, toyoizumi_hamiltonian_2023}. There is still a need to cleanly state how well any QSP-processing methods can succeed for the real polynomials used in the quantum algorithms literature, and whether QSP-processing for complex polynomials can compete against the now standard angle-QSP approaches.

We answer this second question in the affirmative, providing a successful method for complex polynomials. To our knowledge, we provide the largest complex QSP-processing solution for Hamiltonian simulation. Our benchmark cases are standard to the literature: functional approximations used for matrix inversion, Hamiltonian simulation, the eigenvalue threshold problem, and matrix inversion. Our real-valued benchmarks (eigenvalue threshold, matrix inversion) encounter an early difficulty - the coefficient lists of benchmark polynomials are outside the capacities of floating point arithmetic for realistic instances and are beyond our analysis. Our explicit characterization of the affected parameter regimes for rectangle function and inverse function expansions is novel and applies to all known QSP-processing methods using these functional approximations.

\section{QSP Preliminaries}
We review QSP using the framework and notation introduced in \cite{low_hamiltonian_2019, Haah2019product}. We believe this provides the clearest physical intuition for QSP; technical details and proofs of other QSP conventions will diverge slightly from this presentation.

Given a 1-qubit unitary operator $U$ with eigenvalues $e^{i\theta}$ and eigenvectors $\ket{\theta}$, and some basis set in $SU(2)$, $\{\ket{p},\bra{q}\}$, a controlled-$U$ operation \textit{controlled by an ancilla qubit with respect to the $p, q$ basis } is
\begin{equation}
    C_{p}U=\ket{p}\bra{p}\otimes I+\ket{q}\bra{q}\otimes U=VC_0UV^{\dag}.
\end{equation}
 $C_1U$ is the $1$-controlled U gate and $V\in SU(2)$ is the unitary transformation such that $V\ket{0}=\ket{p}$, $V\ket{1}=\ket{q}$.
Acting on an initial state $\ket{\psi_0}=\ket{0}\otimes \sum c_k\ket{\theta_k}$, where $\sum c_k^2=1$,
\begin{align}
    C_pU\ket{\psi_0}&=\sum_kc_k\left(\ket{p}\bra{p}+e^{i\theta}\ket{q}\bra{q}\otimes U\right)\ket{0}\ket{\theta_k}\\
    &=\sum_kc_ke^{i\theta/2}\left(e^{-i\theta/2}\ket{p}\bra{p}\right.\nonumber\\
    &\left.\quad \quad + e^{i\theta/2}\ket{q}\bra{q}\otimes U\right)\ket{0}\ket{\theta_k}.
\end{align}
Assuming we also have access to the inverse operation $C_pU^{\dag}$, then consecutive $C_{p}U, C_{p'}U^{\dag}$ operations will result in the state
\begin{multline}
    \sum_kc_k\left(e^{-i\theta/2}\ket{p}\bra{p}+e^{i\theta/2}\ket{q}\bra{q}\right) \\ \cdot\left(e^{i\theta/2}\ket{p'}\bra{p'}+e^{-i\theta/2}\ket{q'}\bra{q'}\right)\ket{\theta_k},
\end{multline}
where global phases $e^{i\theta/2}, e^{-i\theta/2}$ have cancelled out. Then, a sequence of controlled operations $C_{p_j}UC_{p_{j+1}}U^{\dag}$ is equivalent to the following sequence of operations $E_{p_j}E_{p_{j+1}}U$ on the ancillary
\begin{align}
    F(e^{i\theta/2})&=E_0E_{p_1}(e^{i\theta/2})E_{p_2}(e^{i\theta/2})...E_{p_{2n}}(e^{i\theta/2}),\\
    E_p(e^{i\theta/2})&=e^{-i\theta/2}\ket{p}\bra{p}+e^{i\theta/2}\ket{q}\bra{q}.
\end{align}
$F(e^{i\theta/2})\in SU(2)$ is a polynomial with coefficients determined by products of $\braket{p_k|p_{k+1}}, \braket{q_{k}|q_{k+1}},...$, so it has some Pauli decomposition
\begin{equation}
  F(e^{i\theta})=\mathcal{A}(e^{i\theta})\mathcal{I}+i\mathcal{B}(e^{i\theta})\sigma_X+i\mathcal{C}(e^{i\theta})\sigma_Y+i\mathcal{D}(e^{i\theta})\sigma_Z.
\end{equation}
Laurent polynomials $\mathcal{A, B, C, D}$ are fully determined by projector sets $\{\ket{p_k}\bra{p_k}\}$, are degree at most $n$, are real and have definite reciprocity on the unit circle, and satisfy 
\begin{equation}\label{eq:Haah_sqr_sum_constraint}
    \mathcal{A}^2(e^{i\theta})+\mathcal{B}^2(e^{i\theta})+\mathcal{C}^2(e^{i\theta})+\mathcal{D}^2(e^{i\theta})=1
\end{equation}
for all $\theta\in \mathbb{R}$. Once we measure the ancillary, then some combination of  $\mathcal{A, B, C, D}$ will be prepared on the rest of the circuit. For example, if we measure the ancilla qubit in the Pauli-X basis, then the remainder of the circuit will be prepared in
\begin{equation}
    \ket{\psi_{QSP}}=\sum_kc_k\left(\mathcal{A}(e^{i\theta})+i\mathcal{B}(e^{i\theta})\right)\ket{\theta_k}.
\end{equation}
This description is agnostic to the dimension of $U$, and so we can immediately define the quantum signal processing operator \cref{eq:QET_operator} formalized in \cref{lemma:QSP_Haah}
\begin{align}\label{eq:QET_operator}
U_{QSP}&=\sum_{\theta}\mathcal{P}(\theta)\ket{\theta}\bra{\theta}\\
&=\bra{\cdot}E_0\prod_{k=1}^{n} C_{p_k}UC_{p_{k+1}}U^{\dag}\ket{\cdot},
\end{align}
where $\bra{\cdot}\cdot \ket{\cdot}$ denotes post-selection on some basis measurement on the ancillary, usually $\bra{0}\cdot \ket{0}$ or $\bra{+}\cdot \ket{+}$.
The \textit{target polynomial} $\mathcal{P}$ is a (complex) linear combination of $\{\mathcal{A, B, C, D}\}$ formed by a measurement on the ancillary qubit. Usually we are interested in $\mathcal{P}(z)=\mathcal{A}(z)+i\mathcal{B}(z)$ where $\mathcal{A}$ is reciprocal and $\mathcal{B}$ is anti-reciprocal. See \cref{sec:poly_approx} for a review of Laurent polynomials.

So far, we have shown that a series of controlled operations and transformations $\{V_p\}$ prepares a complex Laurent polynomial of the eigenvalues of $U$. With a complete set of polynomials $\{\mathcal{A, B, C, D}\}$, constructing $\{\ket{p_j}\bra{p_j}, \ket{q_j}\bra{q_j}\}$ is straightforward. Define the matrix-valued coefficient list for polynomial $F^{2n}(e^{i\theta/2})=\sum_{k=-2n}^{2n} C_{k}^{2n}e^{ik\theta/2}$, where the superscript denotes the highest degree with respect to $e^{i\theta/2}$. Then recursively define 
\begin{align}
     \ket{p_j}\bra{p_j}&=\frac{\left(C^{j}_{-n}\right)^{\dag}C^{j}_{-n}}{\text{Tr}\left(\left(C_{-n}\right)^{\dag}C_{-n}\right)}\\
     &\ket{q_j}\bra{q_j}=I-\ket{p_j}\bra{p_j}\\
     F^{j-1}(e^{i\theta/2})&=F^{j}(e^{i\theta/2})E_j(e^{i\theta/2}).
\end{align}
Lemma \ref{lemma:QSP_Haah} summarizes this into the Laurent-QSP convention. For concreteness, the circuit for QSP is given in \cref{fig:QET_basic_circuit}. Note that in practice, we cannot obtain $\{E\}$ exactly, and instead want to identify a suitable set of $\{E\}$ which is some $\epsilon_{qsp}$-close to $\mathcal{P}(e^{i\theta})$. 
\begin{lemma}[Laurent QSP, adapted from \cite{Haah2019product}]\label{lemma:QSP_Haah}
    Consider $2\pi$-periodic function $\mathcal{P}(e^{i\theta})=\mathcal{A}(e^{i\theta})+i\mathcal{B}(e^{i\theta})$. Assume $\mathcal{A}(e^{i\theta})^2+\mathcal{B}(e^{i\theta})^2\leq1$ and $\mathcal{A, B}$ are real-on-circle, pure polynomials. Then, there exists a unique decomposition into $E_p$'s and a unitary 
    $F(e^{i\theta/2})=E_0E_{p_1}(e^{i\theta/2})...E_{p_{2n}}(e^{i\theta/2})$ such that 
    \begin{equation}
        \mathcal{P}(e^{i\theta})=\bra{+}E_0E_{p_1}(e^{i\theta/2})...E_{p_{2n}}(e^{i\theta/2})\ket{+}.
    \end{equation}
\end{lemma}
A central difficulty of QSP in practice is deriving complimentary polynomials. Usually, QSP is constructed beginning from target polynomial $\mathcal{P}(\theta)=\mathcal{A}(\theta)+i\mathcal{B}(\theta)$, and defining complementary polynomial  $i\sqrt{1-\cos^2\theta}\mathcal{Q}(\theta)=-\mathcal{C}(\theta)+i\mathcal{D}(\theta)$. To ensure $\mathcal{C, D}$ fit the requirement \cref{eq:Haah_sqr_sum_constraint}, we have to solve \cref{prob:qsp_poly_constraint}. QSP conventions have historically managed a trade-off between the restrictions on allowed polynomials $\mathcal{P}(\cdot)$ and the conceptual and computational ease of identifying a complementary polynomial $\mathcal{Q}(\cdot)$ in \cref{prob:qsp_poly_constraint}. 
We have phrased the problem generically to stress that \textit{all} QSP conventions have such a step, and the associated polynomial properties restrict available solution methods. 
\begin{problem}[QSP-completion Problem]\label{prob:qsp_poly_constraint}
    Beginning with some polynomial $P$, given as a coefficient list, we want to identify some $Q$ such 
    \begin{equation}
        \left|Q(\cdot)\right|^2=1-\left|P(\cdot)\right|^2
    \end{equation}
    Here  $P(\cdot)$ denotes a polynomial in $z, x,$ or $\theta$
\end{problem}
 Historically, the QSP-completion step is solved with a root-finding routine or with optimization methods. As we show in \cref{sec:fejer}, one can instead use the Fejer problem to solve it.
 
\subsection{QSP Polynomials}
QSP-polynomials are often truncated forms of analytically bounded expansions - for example, the Jacobi-Anger expansion or a trigonometric series. Alternately, \cite{dong_efficient_2021} uses an optimization strategy known as the Remez algorithm to find suitable real polynomials. There is usually a choice of variable - such expansions can be represented in powers of $z=e^{i\theta}, x=\cos\theta,$ or in orders of Chebyshev functions $T_n(x)$. This flexibility is key to understanding the breadth of QSP conventions but can obscure the comparability of each method.

We assume the most general polynomial to implement in QSP has the form $\mathcal{P}: U(1)\rightarrow \mathbb{C}$, such that $|\mathcal{P}(z)|\leq 1$. Using the standard conversion valid on $U(1)$, $z=e^{i\theta}\leftrightarrow x=\cos\theta$,
 \begin{align}
     {P}(z)&=\sum_{k=0}^n a_kz^k=\sum_k^n a_k\left(x+i\sqrt{1-x^2}\right)^k\nonumber\\
     &=\sum_k^n\Tilde{a}_kx^k={P}'(x)
 \end{align}
 for coefficients $a_k, \Tilde{a}_k\in \mathbb{C}$.  Interested readers can refer to \cite{tang_cs_2023} for an overview of Chebyshev combinations for polynomial approximations in QSVT.
 Decomposed into even/odd real-on-circle polynomials,
 \begin{align}
     {P}'(x)&=f_{R,E}(x)+f_{R, O}(x)+if_{I,E}(x)+if_{I, O}(x)
 \end{align}
We emphasize these otherwise pedantic polynomial transformations because it will allow us to relate QSP in every framework. Because $|{P}'|=\sum_i|f_i|^2\leq 1\Rightarrow |f_i|\leq 1$, for $i\in\{(R, E), (R, O), (I, E), (I, O)\}$, each $f_i$ can be implemented in angle-QSP. 

Historically, the strongest methods for solving QSP-processing applied only to $f_{real}:[-1, 1]\rightarrow \mathbb{R}$ \cite{dong_efficient_2021, gilyen_quantum_2019}. In this case, QSP-processing finds a complex function $P_c(x)=f_{real}(x)+if_{imag}(x)$ and a $Q(x)$ (which can be assumed real \cite{dong_efficient_2021}) and then obtains $f_{real}=\frac{1}{2}(P_c(x)+P_c^{*}(x))$ by adding two QSP circuits using a linear combination of unitaries technique. 

This provides a reasonable strategy to implement a generic complex polynomial $P(x)\rightarrow \mathbb{C}$. One can solve QSP-processing for $f_{R,E}, f_{I,E}, f_{R,O}, f_{I,O}$, and then use a linear combination of unitaries (LCU) technique to add the approximations on a circuit. The LCU does not directly increase the query complexity but does introduce a subnormalization on $P$. Many applications will require quantum amplitude amplification to remove this subnormalization, which increases the query complexity to $U$ by a small multiplicative factor.
  
\section{QSP-processing methods}
QSP-processing usually consists of two subroutines; first, a 'completion step,' which consists of identifying $\mathcal{C, D}$ such that \cref{prob:qsp_poly_constraint} is satisfied, and then a 'decomposition step' which works recursively to obtain a set of projectors \cite{Haah2019product} or rotation angles defining the unitary QSP circuit \cite{chao2020finding, gilyen_quantum_2019}. "Direct methods" for QSP-processing solve these steps separately, using root-finding subroutines for the completion step \cite{Ying2022stablefactorization, Haah2019product} and then either a recursive decomposition \cite{Haah2019product} or a least squares solution \cite{chao2020finding} to identify $\{V\}$ in the decomposition step.

Optimization methods for real-valued target functions can compress the completion and decomposition steps, defining a cost function from the QSP circuit and then minimizing its difference from the target function. When the $l$-infinity norm of either the target function \cite{wang_energy_2022} or the coefficients \cite{dong_infinite_2022} obeys stricter bounds, these methods provably converge. However, they are often used successfully outside these conditions.

Optimization for the complex-valued polynomials as used in G-QSP has been shown successful for very high-degree random polynomials; however has not been benchmarked on realistic target functions. \cite{yamamoto2024robust} extends the Prony method technique from \cite{Ying2022stablefactorization} to G-QSP. Their published benchmark is Hamiltonian simulation for $\tau=10$ and $\epsilon_{qsp}=10^{-13}$, computing polynomials with degree up to $\mathcal{O}(10^2)$. 

To summarize, there exist some strong techniques for solving QSP-processing for real-valued functions, using either optimization or direct methods, which under particular conditions are provably convergent. These also represent the most comprehensive benchmarks. It has been an open question whether QSP-processing for complex functions (either G-QSP or Laurent-QSP) can be successful for relevant benchmark polynomials; our contribution is to argue in the affirmative for Laurent-QSP.

\section{Fejer QSP-Processing}\label{sec:fejer}
\subsection{The Fejer Problem}
A central difficulty in QSP-processing is solving \cref{prob:qsp_poly_constraint}, which in the Laurent-QSP convention of \cref{lemma:QSP_Haah}, can be formulated as a known problem in complex analysis.
\begin{definition}[Fejer Factorization, from \cite{goodman_spectral_1997}]
Consider Laurent polynomial $\mathcal{F}(z)=\sum_{k=-n}^nF_kz^k$ such that the coefficients satisfy $F_i=F^*_i=F_{-i}$, and $\mathcal{F}(e^{i\theta})\geq 0$ for $\theta\in\mathbb{R}$. Find a sequence of real numbers $\{\gamma_0, \gamma_1, ....\gamma_n\}$ defining polynomial $\gamma(z)=\sum_{i=0}^{n}\gamma_iz^i$, $\gamma(0)>0$, such that
\begin{equation}
    \mathcal{F}(z)=\gamma(z)\gamma(1/z)
\end{equation}
and $\gamma$ has all roots outside $U(1)$.
\end{definition}
Constraining our solution to have all roots greater than unity and requiring $\gamma(0)>1$ guarantees a unique solution for $\gamma$. Reference \cite{goodman_spectral_1997} outlines several methods for solving the problem and found that two methods with bounded error outperformed root-finding on their benchmarks. Motivated by this early work, we used a quadratically convergent algorithm original to \cite{wilson_factorization_1969} to solve the Fejer problem. 

Our QSP-processing algorithm is algorithm 1 and consists of replacing the root-finding subroutine in \cite{Haah2019product} with the Wilson method given in \cref{sec:wilson}.  Additionally, \cite{Haah2019product} introduces arbitrary precision arithmetic to bind the error and numeric instabilities in the completion and decomposition steps. Herein, we are instead concerned with characterizing the error arising from a straightforward floating point arithmetic implementation on practical polynomials. {Companion paper \cite{skeltonhitchhicker2024} provides a bound-error analysis for algorithm 1.}

\begin{algorithm}
\label{alg:fejer_qsp}
\begin{algorithmic}
\Procedure{QSP-Processing}{$a,b, n, \epsilon_{QSP}$}
\Comment Coefficient lists $a, b$, degree $n$, solution precision $\epsilon_{qsp}$
\If { $a, b$ take arguments on $x\in \left[-1, 1\right]$,} 
  \State convert to length $2n+1$ Laurent polynomial lists $\mathcal{A, B}$.
\EndIf
\If {$a=0$ or $b=0$} 
   \State redefine it as a degree $n-1$ sum of Chebyshev polynomials.
\EndIf
\State Build coefficient list for $1-\mathcal{A}^2(z)-\mathcal{B}^2(z)$ 
\State Solve the Fejer problem for $1-\mathcal{A}^2(z)-\mathcal{B}^2(z)$ using the Wilson method in  \cref{sec:fejer}, obtaining coefficient list $\gamma$
\State Build coefficient lists for $\mathcal{C, D}$ from  $\gamma$
\State Build matrix-valued coefficient list for $$F(z)=\mathcal{A}(z)I+i\mathcal{B}(z)\sigma_X+i\mathcal{C}(z)\sigma_Y+i\mathcal{D}(z)\sigma_Z$$
\State Decompose $F$ using techniques from \cite{Haah2019product}, obtaining $E_0, \{V\}$.
\State \textbf{Return} $E_0, \{V\}$
\EndProcedure
\end{algorithmic}
\caption{QSP-processing with Fejer Methods}
\end{algorithm}

\subsection{Wilson Method}\label{sec:wilson}
We assume access to the coefficients of $\mathcal{F}$, $\{F_i\}$, and an acceptable precision $\epsilon_{fejer}$. Coefficients of $\mathcal{F}$ obey
\begin{equation}\label{eq:wilson_coeff_sum}
   F_i=\sum_{j=0}^{n-i}\gamma_j\gamma_{j+i}.
\end{equation}
Then, we can construct a system of equations for the coefficients that will uniquely specify $\gamma$, and since $F_i=F_{-i}$ for all $i$, we have only $n+1$ distinct coefficients. This can be solved with Newton-Raphson iteration on function $f^l(\gamma^l)=\sum_{j=0}^{n-i}\gamma^l_j\gamma^l_{j+i}-F_l$ and a suitable initial guess $\gamma^{0}$.  First construct vectors 
\begin{equation}
\vec{F}=\begin{bmatrix}
    F_0\\
    F_1\\
    F_2\\
    \vdots\\
    F_n
\end{bmatrix}    \quad
\vec{\gamma}=\begin{bmatrix}
    \gamma_0\\
    \gamma_1\\
    \gamma_2\\
    \vdots\\
    \gamma_n
\end{bmatrix}   
\end{equation}
$\partial_kF_i=\gamma_{k+i}+\gamma_{k-i}$, so define triangular matrices 
\begin{equation}
    T_1=[\gamma_{i+j}]_{ij} \quad T_2=[\gamma_{j-i}]_{ij} \quad  i, j=0,1...n
\end{equation}
 where for $ i+j>n, \:[\gamma_{i+j}]_{ij}=0.$ It is convenient to define the vector form of $ \sum_{j=0}^{n-i}\gamma^l_j\gamma^l_{j+i}$,
 \begin{equation}
    \Vec{c}^l=T_1^l\gamma^l=T^l_2\gamma^l.
 \end{equation}
Then Newton-Raphson iteration step is equivalent to solving the following system of linear equations
\begin{equation}
    (T_1+T_2)\gamma=\Vec{c}+\vec{F},
\end{equation}
which can be solved with standard linear solvers - in our case, NumPy's 'linalg.solve' function. We identify  $\gamma^{k+1}=\frac{1}{2}\gamma^k+\delta^k$ at each step of the NR iteration\footnote{We could take $\delta$ instead, \cite{goodman_spectral_1997} just claims the above expression performs better.}, and ultimately identify a unique choice of $\gamma$ within $\epsilon_{fejer}$. 

Requiring the initial guess $\gamma^0$ to have all zeros outside of the unit circle guarantees that all subsequent $\gamma^k$'s have all zeros outside the unit circle \cite{goodman_spectral_1997, wilson_factorization_1969}. One suitable initial guess is to use constant function $\gamma^{0}(z)=\gamma_0^0$ for some $\gamma^0_0>0$ \cite{wilson_factorization_1969}. 

As summarized in \cite{goodman_spectral_1997}, the Wilson algorithm will converge to a solution in $\mathcal{O}(n^3/3)$ floating point operations.

\section{Benchmarking Methodology}
We benchmark polynomial approximations for $1/x$, $e^{i\tau x}$, and ${rect}(x)$, alongside Fourier series with randomly generated coefficients and a polynomial used in eigenvalue threshold problems; a fuller description of each is presented in \cref{sec:hs_poly}-\cref{sec:matrix_inversion}. Together, these represent a reasonable benchmark set; it contains commonly used approximations in QSP literature and includes both real and complex-valued target functions.
We will evaluate implementations with the following criteria. 
\begin{enumerate}
     \item Is the implementation \textit{successful for noteworthy solution ranges}? That is, does the method provide solutions for the most commonly used functions approximated with QSP, with high approximation precision ($\epsilon_{approx}$) and realistic parameter regimes? For most polynomials, this will require degrees within $10^3, 10^4$.
    \item Relatedly, does the \textit{ implementation precision scale well with polynomial degree}?  We measure the precision of the QSP-processing method as the $l$-$\infty$ norm of the distance between the target polynomial and the QSP-solution, $\epsilon_{qsp}$. We consider $\epsilon_{qsp}\in \left[10^{-14}, 10^{-12}\right]$ to be an acceptably precise result, which is in line with previous benchmarking \cite{Ying2022stablefactorization, dong_efficient_2021}.
    \item Is the implementation \textit{reasonably timed}. We evaluate the number of Newton-Raphson iteration steps instead of the computational time. 
\end{enumerate}
 
\section{Results}
All QSP-processing is done in python\footnote{Code repository is available at \url{https://github.com/Skeltonse/mostly-harmless-QCE24.git}}, although some polynomial coefficients are generated in Julia or Mathematica. We time the completion step only and demonstrate that our technique successfully solves the step up to realistic instance sizes for complex polynomials. Although the Fejer solution can sometimes succeed to $10^{-16}$, input components $\mathcal{A, B}$ can have some error, which $\mathcal{C, D}$ will inherit, as well as error from the Fejer method. Then the input for the decomposition step is only unitary up to  $\epsilon_{coeff}$, and we compute the $l_{\infty}$ norm of the difference between our QSP-expectation value and the polynomial approximation over $[-1, 1]$. We find that using only floating point precision, the error in the decomposition scales polynomially with $n$, as expected.

Recall that all direct methods struggle with this problem. For real-valued polynomials, the halving technique of \cite{chao2020finding} can be used to decompose ${F}(z)$. However, for complex polynomials, there is no known alternative to an iterative decomposition requiring arbitrary precision arithmetic to succeed, as argued by \cite{Haah2019product}.

\subsection{Hamiltonian Simulation} 
\begin{figure}[h]
    \centering
    \includegraphics[width=\linewidth]{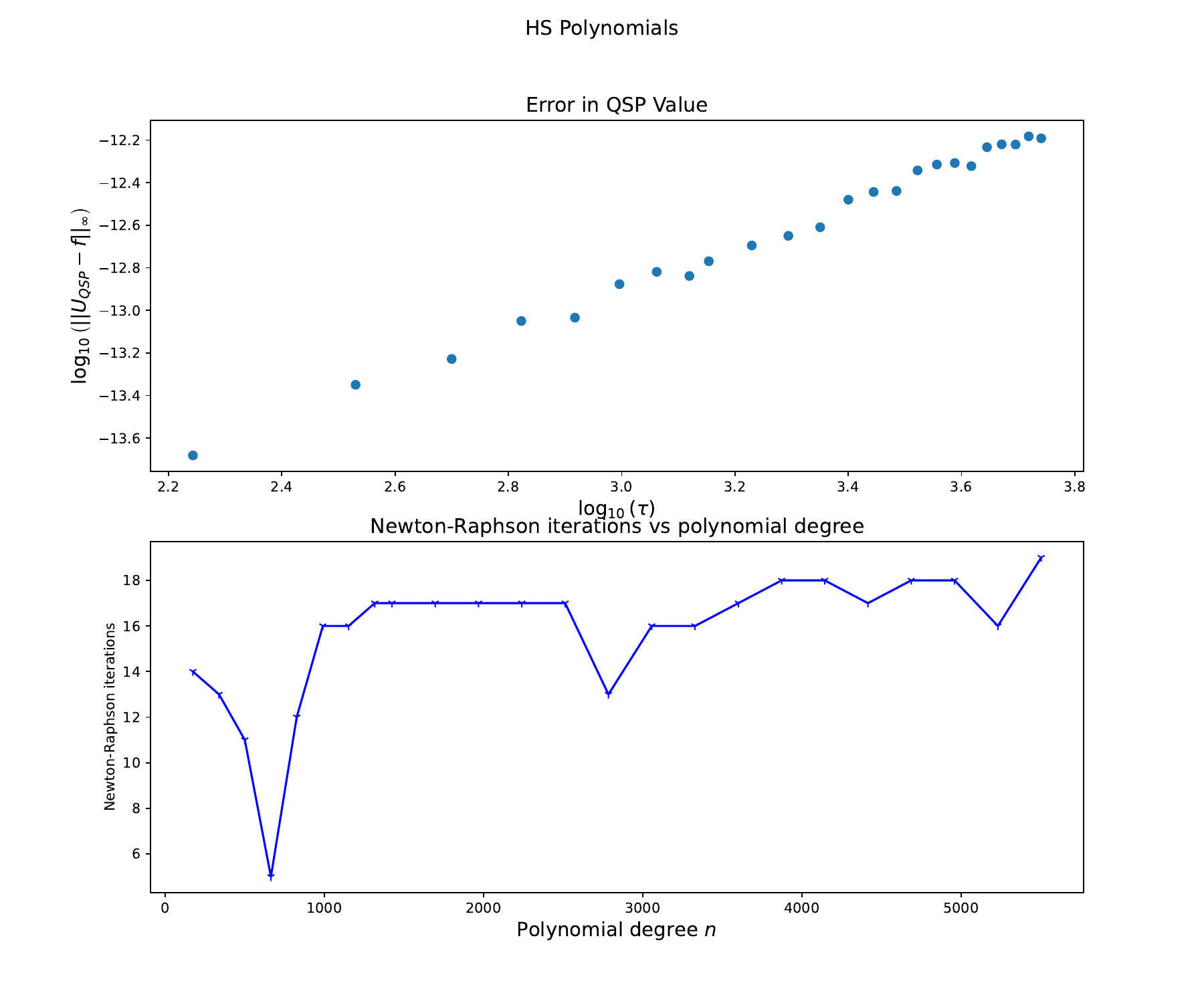}
    \caption{Solution quality for $P_{HS}$ instances within $\tau=20,...1000$. Above: the log-log plot of solution decomposition error against evolution time. Below: completion step times against evolution time.}
    \label{fig:hs_times_its}
\end{figure}

Our target polynomial is found in \cref{sec:hs_poly}. Given precision $\epsilon_{approx}=10^{-14}$, we solve instances with evolution times within $\tau=200, 1000$. \cref{fig:hs_times_its} shows the resulting simulation times. Our completion times appear longer than comparable instances from \cite{dong_efficient_2021, Ying2022stablefactorization}, however, our truncation degree, required for a provably $\epsilon$-bound approximation, leads to longer coefficient lists compared to \cite{dong_efficient_2021}\footnote{For this reason, instances $\tau=4000, 5000$ are outside the space limits of a laptop with machine precision with our truncation, although they are solvable with our method when the truncation proposed in \cite{dong_efficient_2021} is used.}

\subsection{Random Polynomials}
\begin{figure}[h]
    \centering
    \includegraphics[width=\linewidth]{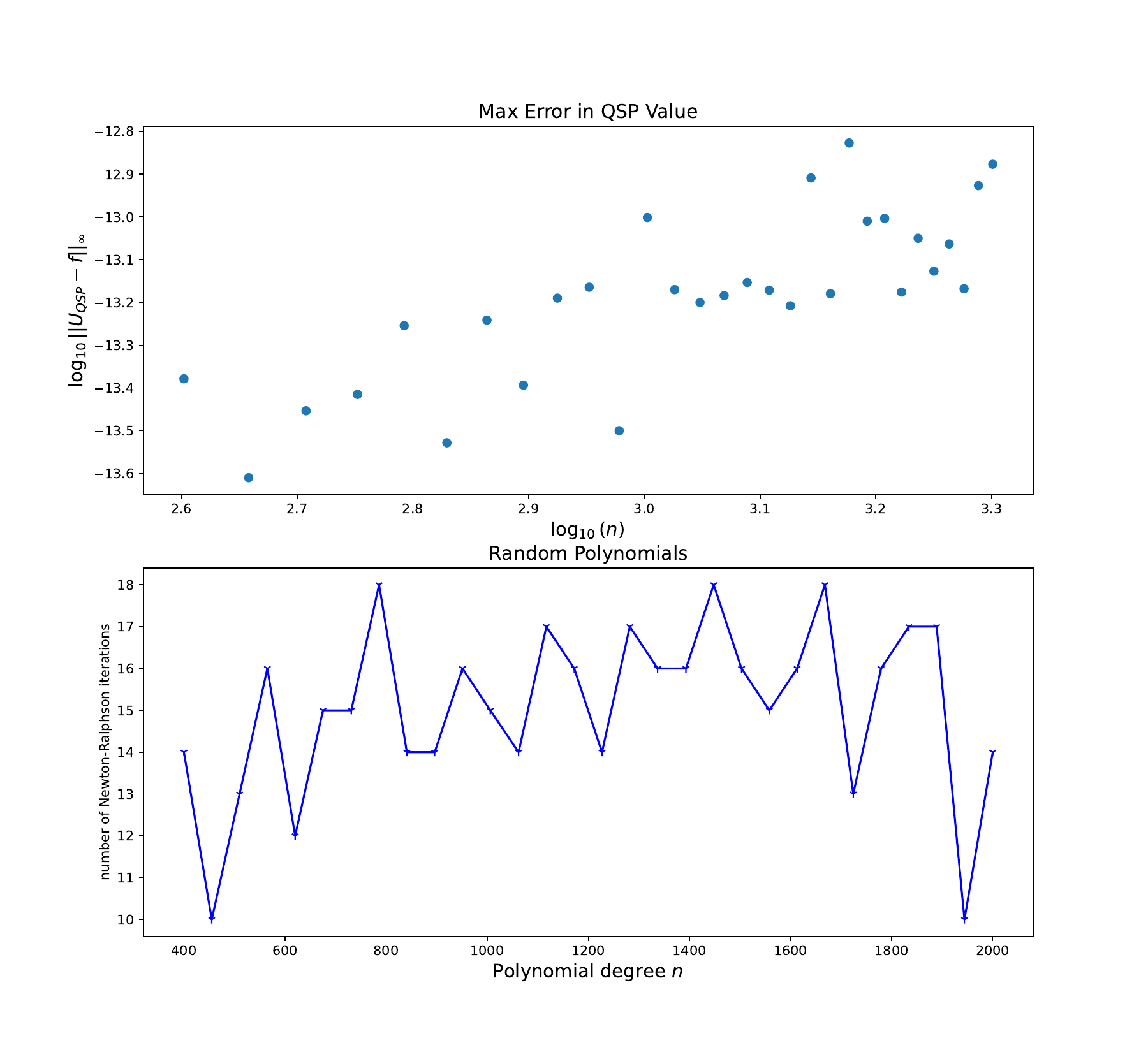}
    \caption{Solution quality for random polynomials with degrees within $\left[400, 2000\right]$ and $\epsilon=10^{-12}$. Above: The log-log plot of solution error against degree demonstrates that the error in our method scales polynomially with the degree. Below: number of iterations scales subexponentially with polynomial degree.}
    \label{fig:random_data}
\end{figure}

For the randomly generated polynomials described in \cref{sec:random_poly}, we find that the solution error and solution time both scale polynomially with increasing degree, see \cref{fig:random_data}. We remark that the barrier to solving polynomials was not polynomial degrees but instead the allowed number of nonzero coefficients. This further reinforces that the practical implementation of QSP-processing is as dependent on the polynomial approximation as on the solution strategy. 

\subsection{Eigenvalue Threshold Problem}
\begin{figure}[h]
    \centering
\includegraphics[width=\linewidth]{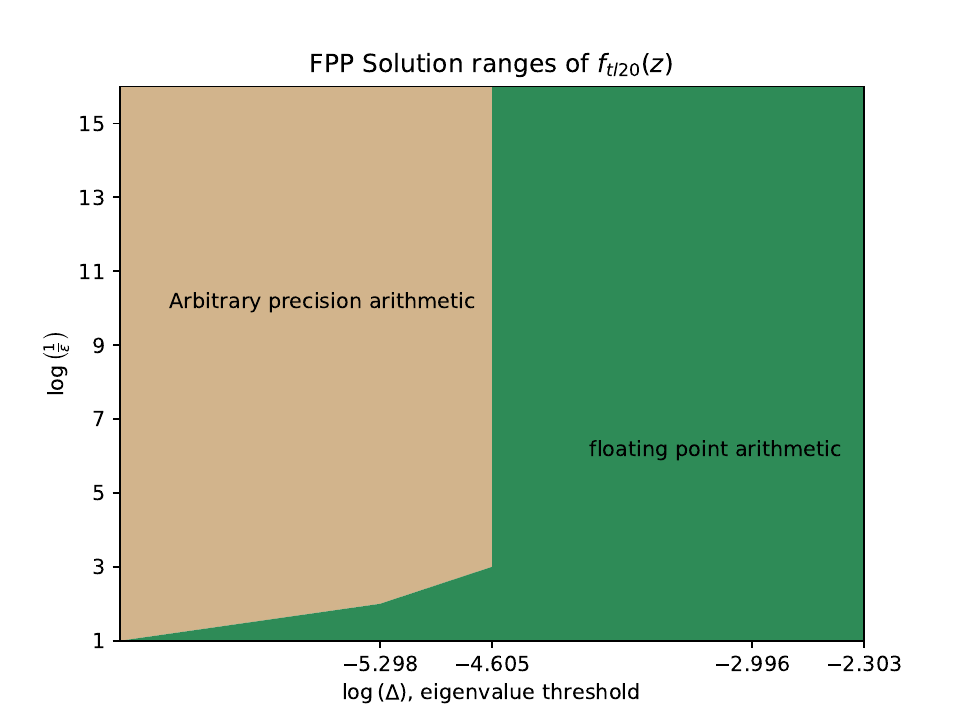}
    \caption{Accessibility of $P_{\textrm{thresh}}(z)$ with floating point arithmetic on a log scale. $\log(1/\epsilon)=1,2...14$, $\Delta\in\{0.1, 0.05, 0.01, 0.05, 0.001\}$, and so most of the interesting applications lie outside of the range of FPP}
    \label{fig:lt_contour_plot}
\end{figure}

For the eigenvalue threshold polynomial defined in \cref{sec:eig_thresh_poly}, we consider spectral gaps  $\Delta\in\{0.1, 0.05, 0.01, 0.05, 0.001\}$ - a similar range to \cite{dong_efficient_2021, Ying2022stablefactorization}. However, the polynomial approximation with respect to $z$ contains extremely large coefficients. Fig. \ref{fig:lt_contour_plot} shows that we cannot generate most instances with high precision (e.g., $10^{-14}$) using machine precision arithmetic in Mathematica. 

Even for instances that can naively be expected to succeed with our method, we find that Mathematica's handling of high-degree Chebyshev polynomials introduces imprecision, which does not allow us to extract coefficients. We note that \cite{dong_efficient_2021} could not solve instances below $\epsilon\approx 10^{-2}$ using direct methods, which aligns with our poor results in this region. For this approximation, the difficulty extracting coefficients is an extremely compelling motivation for working with the Chebyshev basis, where the coefficients have a much simpler form. For this reason, we determine this function is ill-suited to Laurent QSP-processing and do not benchmark it. 

\subsection{Rectangle Function}\label{sec:rect_results}
\begin{figure}[h]
    \centering
\includegraphics[width=\linewidth]{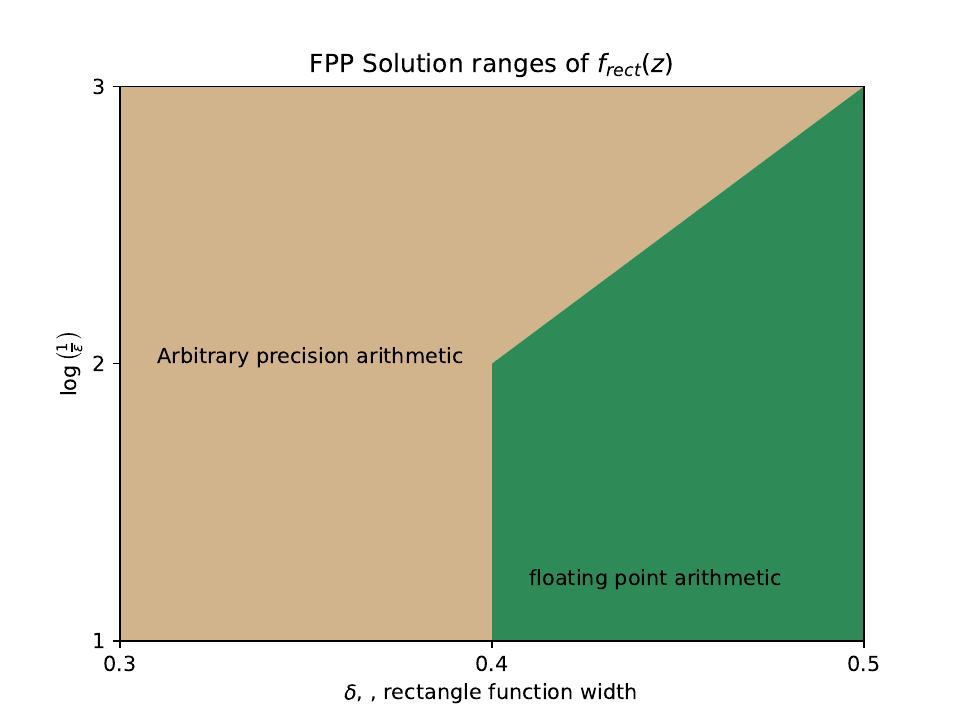}
    \caption{Accessibility of $P_{rect}(z)$ with floating point arithmetic, for $\log(1/\epsilon)=1,2, 3$, $\delta\in\{0.3, 0.4, 0.5\}$}
    \label{fig:rect_contour_plot}
\end{figure}

For the rectangle function defined in \cref{sec:rect_fcn}, we use $\delta\in\{0.3, 0.4, 0.5\}$ and determine when our implementation can compute coefficients within the limits of floating point precision. As shown in \cref{fig:rect_contour_plot}, even for extremely large $\epsilon$, our implementation would require high-precision arithmetic to succeed. Unlike $P_{\textrm{thresh}}$, the Chebyshev basis also contains extremely high coefficients, and thus, this rectangle function approximation is also a challenge to implement in this basis.  

\subsection{Matrix Inversion}
The most well-documented normalization of target polynomials for matrix inversion \cite{gilyen_quantum_2019, martyn_grand_nodate} relies upon the rectangle function in \cref{sec:rect_results} and so all practical instances are out of the range of our methods.
The largest instances in our knowledge were achieved in \cite{dong_efficient_2021} with Matlab, obtaining $\epsilon_{approx}=10^{-14}$,$\kappa=10, 20, ...50$. These instances are still well below the values of $\kappa$, which would be relevant in practical problems. Thus, it is likely that all implementations of quantum algorithms with this function will either use expensive arbitrary arithmetic or simpler approximations for $\frac{1}{\kappa x}$.

The best results given in \cite{Ying2022stablefactorization, dong_efficient_2021} are achieved with simpler approximations, for which $\epsilon$-close analytic bounds to target function $\frac{1}{2\kappa x}$ are not available.

\section{Discussion}
It remains an open question whether solving QSP-processing for interesting problem regimes is as difficult as doing a "double backward somersault through a hoop while whistling the star-spangled banner" \cite{douglas}. Herein, we have approached the choice of target polynomial and its tractability in QSP-processing as intertwined questions whose answers can inform circuit design. In particular, a key question for G-QSP, Laurent-QSP, and a comparable version of multi-variable QSP introduced \cite{németh2023variants} is whether we can follow the success of angle QSP-processing for more general cases.
 
This challenge to G-QSP(s) is especially harsh when one recalls that working with complex-variable QSP may only be advantageous for QSP applications requiring complex approximations. In QSVT, we want to operate on singular values $\chi>0$, where the restriction $\chi\in[-1, 1]$ is fairly benign. However, the exemplar QSP application Hamiltonian Simulation requires precisely this form, and in fact, any Fourier expansion can be represented as a combination of Laurent polynomials. 

We have shown that QSP-processing with the Fejer method can provide high precision and reasonably timed solutions for Laurent polynomials; our benchmarks exemplify different features of QSP-processing polynomials. Importantly, we demonstrated that our completion strategy is successful for high-$\epsilon_{approx}$ instances of Hamiltonian simulation up to $\tau=1000$ and for Fourier series up to degree $10^3$. While the circuit of \cite{motlagh_generalized_2023, berry2024doubling} is optimal for QSP, our decomposition methods succeed for significantly larger $\tau$ instances than given in the G-QSP literature. Additionally, under some conditions, the Fejer method of QSP-processing is provably convergent, see \cite{skeltonhitchhicker2024}.

We have also demonstrated numeric challenges in generating real polynomials for QSP, which sometimes makes the Fejer method unsuitable. For example, the eigenvalue threshold function introduced in \cite{lin_near-optimal_2020} is well-behaved in the Chebyshev basis but is computationally challenging to extract coefficients in the $z$ basis. Contrastingly, all known direct QSP-processing strategies may struggle with sign function approximations in high-$\epsilon_{approx}$ regions, because the coefficients are difficult to represent with floating point arithmetic in any basis. 

This difficulty is not a feature of our QSP-processing method but of the polynomial approximation. It is possible that more sophisticated methods could improve our accessibility plot \cref{fig:rect_contour_plot}. However, our results reinforce that generating coefficient lists, in addition to completing the QSP-processing, is a significant implementation hurdle for QSP. It remains an open question whether any QSP-processing strategy can succeed for high-precision instances of the rectangle or matrix inversion approximations discussed herein. Numeric polynomial approximations, for example, those generated with Remez algorithms, have been more successful  \cite{dong_efficient_2021, Ying2022stablefactorization}. However, they do not seem to skirt problems scaling to high $\epsilon, \kappa$ - recent work \cite{lapworth2024evaluation} used the optimization methods in \cite{dong_efficient_2021} to contain target polynomials, but found that solving practically large instances necessitated larger precision.

Extending our benchmark set to include these types of approximations and explicitly benchmarking against other QSP-processing strategies would be highly interesting future work. Capturing precisely when and where Laurent QSP-processing strategies succeed against angle QSP-processing or G-QSP-processing will be important to determine when a given QSP convention is preferable, especially for intermediate-term algorithms where small multiplicative factors in the circuit depth are meaningful to algorithm success.
 
\section*{Acknowledgment}
The author thanks Tobias Osborne, Anna Kn{\"o}rr, and Ugn{\.e} Liaubaite for helpful discussions. We acknowledge the support of the Natural Sciences and Engineering Research Council of Canada (NSERC), PGS D - 587455 - 2024 and BMWi project ProvideQ.\\

Cette recherche a été financée par le Conseil de recherches en sciences naturelles et en génie du Canada (CRSNG), PGS D - 587455 - 2024.
\bibliographystyle{IEEEtran}
\bibliography{QSVT_bib}

%%%%%%%%%%%APPENDICES%%%%%%%%%%%
\appendix
\section{Target Polynomials}\label{sec:poly_approx}
\subsection{Laurent Polynomials}
We will extensively use Laurent polynomials of degree $n\geq 0$, 
\begin{equation}
    \mathcal{F}(z)=\sum_{k=-n}^nc_kz^k,
\end{equation}
where coefficients $c_k$  can be matrix or scalar-valued. Such polynomials are defined for $\mathbb{C}-\{0\}$, but we will be primarily concerned with their behavior only on $z=e^{i\theta/2}\in U(1)$.
Following \cite{Haah2019product}, we introduce the following terminology
\begin{remark}
    Laurent polynomial $\mathcal{F}(z)$ defined on $z\in \mathbb{C}-\{0\}$ is said to be reciprocal if $\mathcal{F}(z)=\mathcal{F}(z^{-1})$ and anti-reciprocal if $\mathcal{F}(z)=-\mathcal{F}(z^{-1})$. Furthermore, $F(z)$ is
\begin{itemize}
    \item real if all $C_i\in\mathbb{R}$ or $C_i\in \mathbb{M}(\mathbb{R})$ 
    \item real-on-circle if $F(z)\in\mathbb{R}$ for all $z\in U(1)$
    \item pure if it is real-on-circle and reciprocal or if $i\mathcal{F}(z)$ is anti-reciprocal.
\end{itemize}
Note that within the properties reciprocal, real, and real-on-circle, any two of them imply the third.
\end{remark}
An important class of Laurent polynomials for QSP is 
\begin{equation}\label{eq:laur_poly_sdecomp}
    \mathcal{F}(z)=\mathcal{A}(z)+i\mathcal{B}(z),
\end{equation}
defined with real-on-circle Laurent polynomials, $\mathcal{A}(z), \mathcal{B}(z)$ such that $\mathcal{A}(z)=\sum_{l=-n}^n{a}_lz^l$ is reciprocal, and $\mathcal{B}(z)=\sum_{l=-n}^n{b}_lz^l$ is anti-reciprocal.
We will also sometimes require the $l-{\infty}$ norm, denoted $||f||_{\infty}$, and computed as $||f||=\max_{x\in\mathcal{D}}|f(x)|$.

\subsection{Hamiltonian Simulation}\label{sec:hs_poly}
Hamiltonian simulation is one of the most studied QSP/QSVT applications and has become a standard QSP-processing benchmark. The target function for simulation time $\tau$ is $e^{i\tau x}$,  and the Jacobi-Anger expansion is used to construct a target polynomial. The expansion is
\begin{align}
e^{i\tau x}&=\cos(\tau{x})+i\sin(\tau{x})\\
    \cos(\tau x)&=J_0(\tau)+2\sum_{k=1}^{\infty}(-1)^kJ_{2k}(\tau)\cheby_{2k}(x)\\
    \sin(\tau x)&=2\sum_{k=1}^{\infty}(-1)^kJ_{2k+1}(\tau)\cheby_{2k+1}(x),\\
\end{align} 
where $\cheby_j(x)$ is a Chebyshev polynomial of the first kind. We cannot exactly solve for the truncation degree $R$, which bounds precision $\epsilon$ - it would require a solution related to the W-lambeth function. However from \cite{gilyen_quantum_2019}, we can  obtain the following upper bounds

\begin{equation}
        \Tilde{r}(\tau, \epsilon)=\begin{cases}
            \lceil e\tau\rceil, & \tau\geq \frac{\ln(1/\epsilon)}{e}\\
             \lceil \frac{4\ln(1/\epsilon)}{\ln\left(e+\frac{1}{\tau}\ln(1/\epsilon)\right)}\rceil, & \tau\leq \frac{\ln(1/\epsilon)}{e}\\
        \end{cases}
    \end{equation}
    Moreover, for all $q\in\mathbb{R}_+$,
    \begin{equation}
        r(\tau, \epsilon)<e^q\tau+\frac{\ln(1/\epsilon)}{q}.
    \end{equation}

We use
\begin{equation}
        \Tilde{r}(\tau, \epsilon)=\begin{cases}
           e\tau+{\ln(1/\epsilon)} & \tau\geq \frac{\ln(1/\epsilon)}{e}\\
             \lceil \frac{4\ln(1/\epsilon)}{\ln\left(e+\frac{1}{\tau}\ln(1/\epsilon)\right)}\rceil & \tau\leq \frac{\ln(1/\epsilon)}{e}\\
        \end{cases}
    \end{equation}
and then define
\begin{align}
    P_{HS}(z,\tau, \epsilon)&=J_0(\tau)/2+\sum_k^{\Tilde{r}}(-1)^kJ_{2k}(\tau)z^{2k}\nonumber\\
    &+i\sum_k^{\Tilde{r}}(-1)^kJ_{2k+1}(\tau)z^{2k+1}
\end{align}
For use in symmetric-QSP/QSVT, we compute solution sets $\Phi_{\cos}$, $\Phi_{\sin}$ corresponding to complex completions of the $\cos, \sin$ expansions, and use LCU to add the two symmetric-QSP circuits together on the circuit. This introduces a subnormalization of $1/2$, which can be handled using amplitude amplification from \cite{gilyen_quantum_2019}. In this case, Quantum amplitude amplification increases the circuit length by a factor of $3$. 

However, all known QSP-processing methods have introduced an additional subnormalization on $P_{HS}(x,\tau, \epsilon)$ in order to succeed. In practice, we solve QSP-processing with $\frac{1}{2}P_{HS}(x,\tau, \epsilon)$, to degree $2\Tilde{r}$ decompose using Q-QSP to obtain a $4\Tilde{r}$ length circuit and then use any version of amplitude amplification to deal with the subnormalization.

Because $\cos, \sin$ fits the criteria of target polynomials for \cref{lemma:QSP_Haah},  this is an application where we would like Laurent-QSP to offer a speedup. Naively it does, but at a cost of circuit length $2n$. The more efficient decomposition used in G-QSP in \cite{motlagh_generalized_2023, berry2024doubling} can be used to obtain a circuit with query complexity $n+2$ with respect to oracles $CU, CU^{\dag}$ and $U^{\dag}CU^2$ \cite{berry2024doubling}.

\subsection{Random Complex Polynomials}\label{sec:random_poly}
We generate polynomials of the following types:
\begin{align}
    P_{Rand}(x)&=\sum_{j=0}^{n/2}a_{2j}\cheby_{2j}(x)\nonumber\\
    &+i\sum_{j=1}^{n/2}b_j\sin(\arccos(x)) U_{2j}(x)\\
    &=\sum_{j=0}^{n/2}a_{2j}\cos(2j\theta)+i\sum_{j=0}^{n/2}b_{2j+1}\sin((2j+1)\theta)
\end{align}
where wlog we have assumed $n$ is even and  $\chebyt(x)$ is a Chebyshev polynomial of the second kind. This Chebyshev sum is transparently in the form of a Fourier series - such polynomials are popular representations of QSP functions and hence represent a good benchmark. $\{a_j\}, \{b_j\}$ are constructed by sampling over uniform distribution $[0, 1)$; to make the problem more computationally tractable, we keep the total number of coefficients over both polynomials small -  $nz=\min \{\frac{n}{10},40 \}$. Finally, we normalize $\linfnorm{P_{rand}}\leq \frac{1}{2}$, and then it can easily be shown to fulfill the conditions for any QSP procedure. 

\subsection{Eigenvalue Threshold Polynomial}\label{sec:eig_thresh_poly}
We also consider a polynomial used in eigenvalue filtering \cite{lin_optimal_2020}, which is a subroutine for adiabatic quantum linear solvers \cite{lin_optimal_2020} and has been used in quantum algorithms for estimating entropies \cite{wang_new_2022} and fidelity \cite{gilyen_improved_2022}.
The polynomial is  
\begin{align}
    P_{\textrm{thresh}}(x, {\Delta})=\frac{\cheby_k\left(-1 + 2 \frac{x^2-\Delta^2}{1-\Delta^2}\right)}{\cheby_k\left(-1 + 2 \frac{-\Delta^2}{1-\Delta^2}\right)}\\
    k=\lceil\frac{\ln(2/\epsilon)}{\sqrt{2}\Delta}\rceil
\end{align}
and the degree of the polynomial approximation is $n=2k$
From \cite{lin_optimal_2020}, we can relate the action of the approximation is QSP to the action of the projector into the subspace of $H$ corresponding to eigenvalue $\lambda$ with spectral gap $\Delta\in (0, \frac{1}{\sqrt{12}}]$.
\begin{align}
    \left|\left|F^{(SV)}(\Tilde{H}, \Tilde{\Delta})-P_{\lambda}\right|\right|_2\leq e^{2e^{-\sqrt{2}k\Tilde{\Delta}}}\leq \varepsilon\\
    \Rightarrow k\geq \frac{1}{\sqrt{2}\Tilde{\Delta}}\ln(2/\varepsilon)
\end{align}
$\Tilde{H}$ is related to the Hamiltonian of interest as 
\begin{equation}
    \Tilde{H}=\frac{H-\lambda I}{\alpha+|\lambda|},
\end{equation}
and has spectral gap  $\Tilde{\Delta}=\Delta/2\alpha$. 

\subsection{Rectangle Function}\label{sec:rect_fcn}
Symmetrized sign functions are used in QSVT for unstructured search problems, amplitude estimation, and quantum amplitude amplification \cite{martyn_grand_nodate}. A rectangle function is used in the matrix inversion function discussed in \cref{sec:matrix_inversion}. 

We begin with an approximation originally given in thesis \cite{Low_2017} and provide bounds straightforwardly derived from that result. 

\begin{equation}
    \sgn(x-a)=\begin{cases}
        -1 & x<a\\
        1/2 & x=a\\
        1 & x>a\\
    \end{cases}
\end{equation}

We will construct an approximation of the sign function from \cite{low_quantum_2017}, based on the following polynomial approximation of $\erf(kx)$,
    \begin{align}
   p_{\textrm{erf}}(x, k, n)&=\frac{2k e^{-\frac{k^2}{2}}}{\sqrt{\pi}}\left[(-1)^{\frac{n-1}{2}}I_{\frac{n-1}{2}}\left(\frac{k^2}{2}\right)\frac{\cheby_{n}(x)}{n}\right.\\ \nonumber
    &\left.+\sum_{j=0}^{(n-3)/2}(-1)^{j}\left(I_{j}\left(\frac{k^2}{2}\right)\right.\right.\nonumber\\
    &+\left.\left.I_{j+1}\left(\frac{k^2}{2}\right)\right)\cdot\frac{\cheby_{2j+1}(x)}{2j+1}\right].
\end{align}
where $I_j$ is a modified Bessel function of the first kind. $n_{erf}$ is ultimately determined from a bound on the expansion of exponential decay $e^{-\beta(1+x)}$. Given an $\epsilon$-close exponential expansion with truncation degree 
\begin{equation}
    n_{exp}\left(\beta, \epsilon\right)=\lceil\sqrt{2\ln(\frac{4}{\epsilon})\lceil\max\left(\beta e^2, \ln(\frac{2}{\epsilon})\right)\rceil}\rceil,
\end{equation}
the $\epsilon$-close expansion of $erf(kx)$ is truncated at degree
\begin{equation}
    n_{erf}(k, \epsilon)=2n_{exp}\left(\frac{\kappa^2}{2}, \frac{\sqrt{\pi}\epsilon}{4k}\right)+1.
\end{equation}
Then we arrive at
\begin{lemma}[Polynomial Approx to $\sgn$, para. from 6.25-6.27 in \cite{low_quantum_2017}]
    Given $\kappa>0, a\in [-1, 1], \epsilon\in (0, 2\sqrt{2/e\pi})$, the sign function $\sgn(x-a)$ has an $\epsilon$-close approximation on $x\in [-1, a-\kappa/2]\cup [a+\kappa/2, 1]$, given by 
    \begin{align}
   {p}_{\sgn}\left(x, a,\kappa,  \epsilon\right)&=p_{\textrm{erf}}\left(\frac{x-a}{2}, 2k, n\right)\label{eq:low_sign_approx}\\
    n&=2n_{exp}\left(2k^2, \frac{\sqrt{\pi}\epsilon}{16k}\right)+1\\
    &k=\frac{\sqrt{2}}{\kappa}\ln^{1/2}\left(\frac{8}{\pi\epsilon^2}\right) 
\end{align}
\end{lemma}
The rectangle function is defined as  
\begin{align}
rect\left(\frac{x}{2t}\right)&=\begin{cases}
    0, \quad |x|>t\\
    \frac{1}{2}, \quad |x|=t\\
    1, \quad |x|<t\\
\end{cases}\\
&=\frac{1}{2}\left(\sgn(x+t)-\sgn(x-t)\right)
\end{align}
where $t$ is half the width of the interval. We want an $\epsilon$-close approximation using two sign function approximations. The difficulty is that adding two sign approximations means introducing two ranges where the approximation does not succeed. Say we require the approximation to be valid outside of a $\delta$-length range defined by $[-t-\delta/2, -t]\cup [t, t+\delta/2]$. This  corresponds to sign function invalidity ranges $[-a-\frac{\kappa}{2}, -a+\frac{\kappa}{2}], [a-\frac{\kappa}{2}, a+\frac{\kappa}{2}]$.  Then,
\begin{align}
 P_{rect, t, \delta, \epsilon }(x) &=\frac{1}{2}\left(p_{,\kappa, \epsilon }(x+a)-p_{,\kappa, \epsilon }(x-a)\right)\label{eq:rect_approx}\\
    a&=t+\frac{\delta}{4}, \quad \kappa=\frac{\delta}{2}
\end{align}
An entirely straightforward calculation shows that $|rect(x/2t)-p_{rect, t, \delta, \epsilon }(x)|\leq \epsilon$ on the range of validity. Technically, the approximation is bound by $|p_{rect, t, \delta, \epsilon }(x)|\leq 1+\epsilon$, but since our QSP methods require a subnormalization of $1/2$ anyways, we can disregard this.

\subsection{Matrix Inversion}\label{sec:matrix_inversion}
The most common function target function for matrix inversion of $A$ is 
\begin{equation}
    \frac{1-rect({\kappa x})}{2\kappa x}.
\end{equation}
where the rectangle function is introduced to ensure that we can normalize the function over the range $\left[-1/2\kappa, 1/2\kappa\right]$ defined by matrix $A$'s condition number $\kappa$, where our approximation is otherwise greater than $1$. 

For the target polynomial, we use a function introduced for the quantum walk-inspired algorithm for solving a system of linear equations \cite{childs_quantum_2017},
\begin{align}
    P_{1/x}(x)&= 4\sum_{j=0}^{j_0}(-1)^j\left[\frac{\sum_{i=j+1}^b{2b \choose b+i}}{2^{2b}}\mathcal{T}_{2j+1}(x)\right].\\
    b&=\lceil(\kappa)^2\ln(\kappa/\epsilon)\rceil\\
    j_0&=\lceil\sqrt{b\ln(4b/\epsilon)}\rceil
    \label{chebyshev_fcn}
\end{align}
$g(x)$ is $2\epsilon$ close to $\frac{1}{ x}$ on $\mathcal{D}_{\kappa}=[-1, \frac{-1}{\kappa})\cup (\frac{1}{\kappa}, 1]$. 

\cite{gilyen_quantum_2019} extends this algorithm to the QSVT framework - now the sparse-access oracles for $A$ introduced in \cite{berry_hamiltonian_2015} prepare a block encoding of $A$, and instead of using an LCU to add the Chebyshev functions of $A$'s eigenvalues, a QSVT circuit computes function on each singular value of $A^{\dag}$. Like with the original quantum walk framework, the query complexity of implementing this block encoding scales with the sparsity $d$, $\kappa$, and precision $\epsilon$.

However, for QSVT, we must manage the normalization within the functional approximation. 
With $P_{rect, \epsilon', \kappa}$ defined from \cref{eq:rect_approx}, the matrix inversion function is
\begin{align}
    P_{MI, \epsilon, \kappa}=\frac{1}{2\kappa}P_{1/x, \epsilon_{inv}, \kappa}(x)\left(1-P_{rect, \epsilon_{rect}, \frac{1}{2\kappa},\frac{1}{\kappa}}(x)\right)
\end{align}
where $\epsilon_{inv}, \epsilon_{rect}$ are defined by
\begin{align}
    \epsilon_{rect}(\kappa, \epsilon)&=\min\left(\epsilon,\frac{\kappa}{2j_0} \right)\\
    n_{rect}(\kappa, \epsilon_{rect})&=1+2n_{exp}\left(2k^2, \frac{\sqrt{\pi}\epsilon}{8k}\right)\\ 
    k_{rect}&={\sqrt{2}\kappa}\ln^{1/2}\left(\frac{8}{\pi\epsilon^2}\right)\\
    b(\kappa, \epsilon)&=\lceil(\kappa)^2\ln(\kappa/\epsilon)\rceil\\
    n_{inv}&=\lceil\sqrt{b\ln(4b/\epsilon)}\rceil\\
    n_{MI}&=n_{inv}+n_{rect}.
\end{align}
We omit the derivation of these bounds for space; similar proofs are found in \cite{gilyen_quantum_2019, martyn_grand_nodate}. If $A$ is Hermitian, we can use QSP and oracle  $U^{i\arccos(A)}$, quite similarly to what is done in \cite{Haah2019product}. If $A$ is not hermitian, then we must use QSVT.

\section{QSP Circuit}

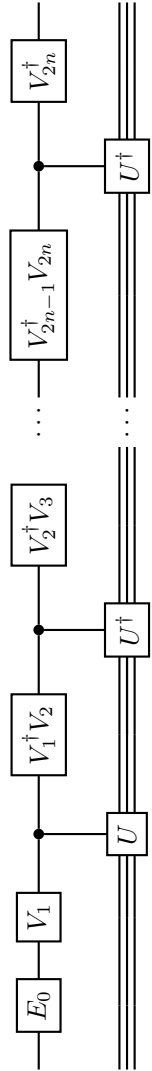
\begin{sidewaysfigure}
    \centering
    \begin{quantikz}
\qw & \gate{E_0} & \gate{V_1} & \ctrl{1} & \gate{V_1^{\dag}V_2} &  \ctrl{1}& \gate{V_2^{\dag}V_3} &\ \ldots\  & \gate{V^{\dag}_{2n-1}V_{2n}} & \ctrl{1}  & \gate{V_{2n}^{\dag}} & \qw\\
\qwbundle[alternate]{} & \qwbundle[alternate]{}  & \qwbundle[alternate]{}& \gate{U}\qwbundle[alternate]{} & \qwbundle[alternate]{} 
& \gate{U^{\dag}}\qwbundle[alternate]{} & \qwbundle[alternate]{} & \qwbundle[alternate]{} \ \ldots\ & \qwbundle[alternate]{}  &  \gate{U^{\dag}}\qwbundle[alternate]{}  &  \qwbundle[alternate]{}&  \qwbundle[alternate]{}
\end{quantikz}
    \caption{Basic QET circuit for operator $U$, preparing  preparing $\mathcal{A}+i\mathcal{B}: \{\lambda\}\rightarrow |z|<1$ given set of unitaries $\{V\}\in SU(2)^{\otimes n}$, $n=2k$.}
    \label{fig:QET_basic_circuit}
\end{sidewaysfigure}

\end{document}